\documentclass[nofootinbib,twocolumn,prl,preprintnumbers]{revtex4-1}
\pdfoutput=1
\usepackage{amsmath,amsthm,amssymb,multirow,psfrag}
\usepackage{epsfig}
\usepackage{color}

\graphicspath{{./Figures/}}

\begin{document}

\def\lsim{\mathrel{\rlap{\lower4pt\hbox{\hskip1pt$\sim$}}
  \raise1pt\hbox{$<$}}}
\def\gsim{\mathrel{\rlap{\lower4pt\hbox{\hskip1pt$\sim$}}
  \raise1pt\hbox{$>$}}}
\newcommand{\vev}[1]{ \left\langle {#1} \right\rangle }
\newcommand{\bra}[1]{ \langle {#1} | }
\newcommand{\ket}[1]{ | {#1} \rangle }
\newcommand{\ev}{ {\rm eV} }
\newcommand{\kev}{{\rm keV}}
\newcommand{\mev}{{\rm MeV}}
\newcommand{\tev}{{\rm TeV}}
\newcommand{\mpl}{$M_{Pl}$}
\newcommand{\mw}{$M_{W}$}
\newcommand{\Ft}{F_{T}}
\newcommand{\Zparity}{\mathbb{Z}_2}
\newcommand{\BLambda}{\boldsymbol{\lambda}}
\newcommand{\met}{\;\not\!\!\!{E}_T}
\newcommand{\beq}{\begin{equation}}
\newcommand{\eeq}{\end{equation}}
\newcommand{\bea}{\begin{eqnarray}}
\newcommand{\eea}{\end{eqnarray}}
\newcommand{\nn}{\nonumber}
\newcommand{\gev}{{\mathrm GeV}}
\newcommand{\hc}{\mathrm{h.c.}}
\newcommand{\eps}{\epsilon}
\newcommand{\bwt}{\begin{widetext}}
\newcommand{\ewt}{\end{widetext}}
\newcommand{\draftnote}[1]{{\bf\color{blue} #1}}

\newcommand{\cO}{{\cal O}}
\newcommand{\cL}{{\cal L}}
\newcommand{\cM}{{\cal M}}

\newcommand{\fref}[1]{Fig.~\ref{fig:#1}} 
\newcommand{\eref}[1]{Eq.~\eqref{eq:#1}} 
\newcommand{\aref}[1]{Appendix~\ref{app:#1}}
\newcommand{\sref}[1]{Section~\ref{sec:#1}}
\newcommand{\tref}[1]{Table~\ref{tab:#1}}

\title{\Large{{\bf Golden Probe of the di$-\Upsilon$ Threshold}}}

\author{{\bf {Yi Chen$\,^{a}$,~Roberto Vega-Morales$\,^{b}$
}}}

\affiliation{
$^a$CERN, European Organization for Nuclear Research, Geneva, Switzerland\\
$^b$Departamento de F\'{i}sica Te\'{o}rica y del Cosmos and CAFPE,~Universidad de Granada, Campus de Fuentenueva, E-18071 Granada, Spain
}
\preprint{UG-FT 324/17, CAFPE 194/17}
\email{
yichen@caltech.edu\\
rvegamorales@ugr.es\\}
\begin{abstract}
Recent studies indicate that the ground state of a QCD four bottom quark system may form a resonance in the range $\sim 18 -19$~GeV, near the di-$\Upsilon$ threshold, which may be observable at the LHC in the four lepton final state.~These studies also predict a variety of possible resonances associated with the various excited states and allow for a number of possible spin and $CP$ quantum numbers.~Of course, even if a resonance is observed in four leptons in the predicted mass range, it will be prudent to experimentally confirm its  four bottom quark QCD nature or whether it is perhaps something more exotic.~We initiate an investigation in this direction by exploring the ability of the \emph{normalized} fully differential decay width for decays to four leptons to probe the underlying nature of a putative resonance including $CP$ and tensor properties.~We assume the first observed state is a spin-0 boson decaying to at least one on-shell $\Upsilon$, but allow for decays to four leptons through $\Upsilon\Upsilon,\,\Upsilon\gamma$,~and $\Upsilon Z$ vector boson pairs.~We consider a range of resonance masses around the di-$\Upsilon$ threshold and find excellent prospects at the LHC for establishing its $CP$ and tensor properties, perhaps not long after a discovery depending on (unknown) production cross sections and the exact resonance mass.
\end{abstract}
\maketitle

\section{Introduction} \label{sec:intro} 

Recent studies indicate~\cite{Berezhnoy:2011xn,Du:2012wp,Chen:2015moa,Karliner:2016zzc,Chen:2016jxd,Bai:2016int,Wu:2016vtq,Richard:2017vry,Wang:2017jtz,Eichten:2017ual} that a QCD bound state composed of four bottom quarks, a so called `beauty-full' tetraquark~\cite{Bai:2016int}, may be observable as a resonance at the LHC in the mass range $\sim 18-19$~GeV\,\footnote{Tantalizingly, a recent (not yet approved) CMS analysis~\cite{CMSpub} hints at a potential four lepton excess just in this mass range.}.~In particular it has been pointed out that decays to an on-shell $\Upsilon$ and a pair of leptons with $\Upsilon$ also decaying to leptons allows for the possibility of a striking signal at the LHC in the invariant mass distribution of $\Upsilon(1S) \,\ell^+\ell^-$ events ($\ell = e,~\mu$).~These studies predict a plethora of bound states around, but just \emph{below}, the di-$\Upsilon, \Upsilon/\eta_b$ thresholds~\cite{Chen:2016jxd}, thus implying a bottom tetraquark bound state.~This includes bound states of spin~$0,1,$~or~$2$ carrying different combinations of $CP$ quantum numbers, potentially including admixtures.~While progress on computing the ground state energy of a tetraquark bound state (or effective mass) has been made,~studies of their decays are still in the preliminary~\cite{Berezhnoy:2011xn,Chen:2016jxd,Karliner:2016zzc,Eichten:2017ual} stages.

These studies of potential tetraquark systems are based on theoretical calculations utilizing various sum rules or \emph{ad hoc} bound state potentials to compute the $4b$ ground state.~Generically these calculations have found the ground state to lie below the di-$\Upsilon$ threshold.~However, recent lattice QCD calculations based on first principles~\cite{Hughes:2017xie} indicate that no $4b$ bound state is expected below the di-$\Upsilon$ threshold.~Instead the ground state is found to be on threshold, implying a di-meson system and not a tetraquark bound state.~Regardless, whether a putative $4b$ resonance appears above or below the di-$\Upsilon$ threshold there is the possibility for a striking four lepton signal which could be observable at the LHC.

The fact that a potentially narrow resonance\,\footnote{We comment that the authors in~\cite{Anwar:2017toa} studied $4b$ tetraquark bound states and found it would \emph{not} have a narrow width.}, can be observed in the precisely measured four lepton channel offers a unique opportunity to study its detailed properties utilizing the differential spectra of the many observables in the final state.~Furthermore, even if a resonance is observed in four leptons in the predicted mass range, it will be prudent to experimentally confirm its  four bottom quark QCD nature or whether it is perhaps something more exotic.~We initiate an investigation\,\footnote{Recently the authors in~\cite{Eichten:2017ual} also examined the $4\ell$ channel as well as discussed potential production mechanisms of a putative $4b$ bound state decaying through an intermediate $\Upsilon\Upsilon^\ast$ pair.}
of this possibility assuming at least one on-shell $\Upsilon$, but making minimal assumptions of how the effective couplings to $\Upsilon \,\ell^+\ell^-$ are generated since these couplings can depend on both perturbative and non-perturbative effects~\cite{Wang:2017jtz} as well as in principle beyond the SM (BSM) physics.

To this end we utilize a generic effective theory framework to analyze these (potential) tetraquark decays.~In particular, we explore the possibility that the `off-shell' lepton pair can come from either an $\Upsilon,~Z$, or $\gamma$ vector boson.~While in the SM we expect the $\Upsilon$ contribution to dominate~\cite{Bai:2016int,Eichten:2017ual}, we examine whether the fully differential decay width for decays to four leptons can be used to establish this directly.~Should the $\gamma$ or $Z$ be found to dominate or contribute significantly, it would be a tantalizing hint that perhaps this is not a SM QCD state and instead something more exotic.~As part of this analysis we explore the possibility of probing the $CP$ and tensor properties.~We also briefly discuss various potential $CP$ violating effects which, should a resonance be discovered, would be interesting to investigate further.

For this simplified study we assume the first observed resonance is a spin-0 scalar allowing for various $CP$ and tensor structures in its couplings to $\Upsilon\Upsilon,\Upsilon Z,\Upsilon\gamma$ pairs.~We utilize the matrix element method (MEM) analysis framework developed in~\cite{Chen:2012jy,Chen:2013ejz,Chen:2014pia,Chen:2014gka,Chen:2014hqs,Falkowski:2014ffa,Chen:2015iha} for scalar decays to four leptons which includes the tree level $q\bar{q} \to 4\ell~(4\ell \equiv 2e2\mu,\,4e,\,4\mu)$ background.~Motivated by the exciting possibility of a tetraquark bound state, we emphasize in our analysis a spin-0 resonance $\sim 500$~MeV \emph{below} the di-$\Upsilon$ threshold, but also examine  resonance masses just below, at, and above the di-$\Upsilon$ threshold.

We perform various hypothesis tests to examine the ability to directly establish the nature of the putative resonance including its $CP$ and tensor properties.~We find excellent prospects once $\mathcal{O}(10 - 100)$ signal events have been observed at the LHC which may in principle be possible not long after discovery depending on (unknown) production cross sections and the exact mass of the putative resonance.~We leave a study of spin-1 and spin-2 resonances decaying to $2\ell + \Upsilon$ as well as other multiquark states~\cite{Karliner:2017qhf} with $4\ell$ signals to future work.

\section{Decays to Four Leptons} \label{sec:fourleps}
Here we discuss the effective theory framework we use to model the decays to four leptons of a putative scalar resonance.~To gain intuition for the discriminating power of the fully differential decay width, we also examine the phase space contributions of the operators in~\eref{LM} and~\eref{LVV} as well as some invariant mass and angular distributions in the $4\ell$ center of mass frame~\cite{Chen:2012jy,Chen:2013ejz}. 

\subsection{Modeling $\varphi \to \Upsilon + \ell^+\ell^- \to 4\ell$ decays} \label{sec:decays}

To minimize any dependence on an unknown production mechanism we will therefore only use `decay observables' in our MEM analysis~\cite{Chen:2013ejz} discussed more below.~Our task is further simplified by assuming an on-shell decay of a narrow resonance and taking its mass as input as well as imposing the experimental constraint~\cite{Khachatryan:2016ydm} that one $\ell^+\ell^-$ pair come from an \emph{on-shell} $\Upsilon$.~As discussed above, the second lepton pair can be due to a $Z$,~$\gamma$,~or $\Upsilon$ vector boson.~We also expect decays into $\eta_b$ scalar mesons and subsequent decay into leptons are suppressed relative to $\Upsilon$ by the required spin flip of the $b$-quark~\cite{Berezhnoy:2011xn} as well as the negligible branching ratio into leptons~\cite{Olive:2016xmw}.~We therefore neglect these contributions to the dilepton spectrum.~We have also neglected contributions from the $\Upsilon(2S)$ and $\Upsilon(3S)$ excited states.~However, including these various contributions may be necessary once many four lepton events have been collected.

Approximating electrons and muons to be massless we can write the tensor structure of a spin-0 particle which couples to two conserved lepton (pair) currents as,
\bea
\label{eq:S0vert}
\Gamma^{\mu\nu}_{0} 
&=& 
\mathcal{C}_1 \, g^{\mu\nu} 
+ \mathcal{C}_2 \, k_1^\nu k_2^\mu + 
\mathcal{C}_3 \, \epsilon^{\mu\nu\alpha\beta}k_{1\alpha} k_{2\beta} ,
\eea
where the $\mathcal{C}_n$ are complex momentum dependent form factors which generically can contain multiple poles.~Since an on-shell $\Upsilon$ is required in each event, it always makes up one of the poles.

The invariant mass of the second lepton pair can be above or below the $\Upsilon$ mass and there may be sizable contributions from off shell photons especially at low dilepton invariant mass near the photon pole.~There is in principle also an (highly) off-shell $Z$ contribution which is expected to be suppressed by the large $Z$ boson mass as compared to four lepton invariant mass.~However, for something more exotic than a SM QCD state, these $Z$ and $\gamma$ contributions could in principle be enhanced by BSM effects.~Thus for our study we assume the decay topology for the spin-0 resonance decaying to four leptons to be $\varphi \to \Upsilon(1S) V \to 4\ell~~(V = \Upsilon, Z, \gamma)$ and parametrize the $\varphi \Upsilon V$ couplings with effective operators imposing only Lorentz and electromagnetic gauge invariance.

At dimension three this gives the `mass operators' involving $\Upsilon\Upsilon$ and $\Upsilon Z$ vector boson pairs,
\bea\label{eq:LM}
\mathcal{L}_{M} 
&\supset& 
m_\varphi \, \varphi \, (
 \mathcal{A}_{1}^{\Upsilon \Upsilon} \, \Upsilon^{\mu}\Upsilon_{\mu} 
+
\mathcal{A}_{1}^{\Upsilon Z} \, \Upsilon^{\mu}Z_{\mu} ) ,
\eea
where $m_\varphi \equiv M_{4\ell}$ is the putative resonance mass and equivalent to the four lepton invariant mass assuming an on-shell $\varphi$ decay.~The form factors $\mathcal{A}_n^{i}$ can in principle be complex, but this will not affect our analysis below and we now neglect any momentum dependence.~The momentum dependence of these form factors may be relevant,~but this requires an analysis of both perturbative and non-perturbative effects which is beyond the scope of this study and left to future work.

Allowing for operators up to dimension five, including both $CP$ odd and $CP$ even, we consider the set,
\bea\label{eq:LVV}
\mathcal{L}_{\Upsilon \Upsilon} 
&\supset& 
\frac{\varphi}{m_\varphi} 
(
\mathcal{A}_{2}^{\Upsilon \Upsilon}  \, \Upsilon_{\mu\nu}\Upsilon^{\mu\nu}
+ 
\mathcal{A}_{3}^{\Upsilon \Upsilon} \, \Upsilon_{\mu\nu} \widetilde{\Upsilon}^{\mu\nu}), \,\nonumber \\
\mathcal{L}_{\Upsilon Z} 
&\supset& 
\frac{\varphi}{m_\varphi} 
(
\mathcal{A}_{2}^{\Upsilon Z} \, \Upsilon_{\mu\nu} Z^{\mu\nu}
+  
\mathcal{A}_{3}^{\Upsilon Z} \, \Upsilon_{\mu\nu} \widetilde{Z}^{\mu\nu}) , \,  \\
\mathcal{L}_{\Upsilon \gamma} 
&\supset& 
\frac{\varphi}{m_\varphi} 
(
\mathcal{A}_{2}^{\Upsilon \gamma} \, \Upsilon_{\mu\nu}F^{\mu\nu}
+ 
\mathcal{A}_{3}^{\Upsilon \gamma} \, \Upsilon_{\mu\nu} \widetilde{F}^{\mu\nu} \,  \nonumber
) ,
\eea
where again we neglect any momentum dependence.~While this list is not exhaustive, including the additional operators does not qualitatively affect our analysis or change our general conclusions.~Note that the $Z$ boson contribution effectively generates dimension five contact operators (e.g.~$\varphi \Upsilon^\mu \bar{\ell} \gamma_\mu \ell $) though a more general analysis would include these separately.~Thus our main assumption is that the operators in~\eref{LM} and~\eref{LVV} capture the relevant momentum structure and poles.

We have implicitly assumed an expansion in $m_\varphi^{-1}$ is valid, but since our analysis uses only \emph{shape} decay information to conduct simple hypothesis tests, our results are independent of any normalization assumed.~This also mitigates any dependence on non-perturbative QCD effects which might affect the overall normalization\,\footnote{We thank Ciaran Hughes for pointing this out.}.~The presence of additional poles appearing in the $\mathcal{A}_n^{i}$ form factors due to either non perturbative or beyond the SM effects could invalidate this expansion as well as generate a complex phase.~Though we do not examine this possibility here since our analysis is independent of whether they are complex, we will discuss it briefly below.

To model the $\Upsilon$ decays to leptons we follow closely the analysis of~\cite{Aloni:2017eny}.~Assuming only SM contributions, we parameterize the $\Upsilon$ couplings to leptons as,
\bea\label{eq:LU}
\mathcal{L} \supset 
(\frac{f_\Upsilon}{m_\Upsilon} A_\Upsilon^{SM}) 
\Upsilon^\mu \bar\ell \gamma_\mu \ell ,
\eea
where $A_\Upsilon^{SM}$ is a dimensionless parameter which depends on both perturbative effects, and on non-perturbative meson-to-vacuum matrix elements.~Since we are assuming only the SM in the $\Upsilon \to 2\ell$ decay we have~\cite{VanRoyen:1967nq} at leading order $A_\Upsilon^{SM} \approx - 4\pi \alpha Q_b$.~There are small corrections~\cite{Aloni:2017eny} to this from various effects, but they can be safely neglected for our purposes.~The form factor $f_\Upsilon$ parameterizes the vacuum to meson transition amplitude and must be determined by measurements or lattice calculations.~For purposes of the present study we simply treat the separate form factors as one effective coupling $g_\Upsilon \equiv (f_\Upsilon A_\Upsilon^{SM}/m_\Upsilon)$ in~\eref{LU}.~We approximate $g_\Upsilon$ numerically from the measured branching ratio of $\Upsilon$ decaying into muons ($\approx 2.5\%$) which gives $|g_ \Upsilon| \approx 0.00232$ assuming $m_\Upsilon = 9.46$~GeV and $\Gamma_\Upsilon = 54\times 10^{-6}$~GeV~\cite{Olive:2016xmw}.~Similarly to the $Z$ boson, the propagation of the intermediate $\Upsilon$s is modeled with a spin-1 massive vector boson propagator of the form,
\bea
\mathcal{P}^\Upsilon_{\mu\nu}
\sim
\frac{g_{\mu\nu} - k_\mu k_\nu/m_\Upsilon^2}{k^2 - m_{\Upsilon}^2 + i m_{\Upsilon}\Gamma_{\Upsilon}},
\eea
where the $k_\mu k_\nu$ term drops out when dotted into the conserved currents associated with each pair of the final state (massless) charged leptons.


\subsection{Fully differential decay width\\
and integrated magnitudes}
\label{subsec:fullydiffw}

The basis for our analysis will be the $\varphi \to 4\ell$ fully differential decay width analytically computed and validated in~\cite{Chen:2012jy,Chen:2013ejz,Khachatryan:2014kca} for Higgs decays, but adapted here to include an intermediate $\Upsilon$.~All interference effects are computed including those from identical final state interference in the case of the $4\mu, 4e$ final state as well as interference between the operators in~\eref{LM} and~\eref{LVV} though they will not be relevant for our results below.

The fully differential decay width for $\varphi\rightarrow 4\ell$ can be written as a sum over terms quadratic in the couplings which we can write schematically as,
\bea
\label{eq:diffwidth}
\frac{d\Gamma_{\varphi\rightarrow 4\ell}}{d\mathcal{P}} 
&\sim& \sum \mathcal{A}_n^{i} \mathcal{A}_{m}^{j\ast} 
\times \frac{d\Gamma_{nm}^{ij}}{d\mathcal{P}},
\eea
where the sum is over $n,m = 1,2,3$ and $i,j =  \Upsilon\Upsilon,  \Upsilon Z, \Upsilon\gamma$ (note $\mathcal{A}_1^{\Upsilon\gamma} = 0$) and we have neglected any momentum dependence in the couplings as discussed above.~Here $d\mathcal{P}$ represents the differential volume element, or phase space.~In the four lepton center of mass frame this can be parametrized in terms of two invariant masses formed out of the four vectors of the di-lepton pairs as well as five angular variables~\cite{Chen:2013ejz,Chen:2014pia}.

In order to gain intuition on the potential to discriminate between the operators in~\eref{LM} and~\eref{LVV} it is useful to examine what we call the \emph{integrated magnitudes}~\cite{Chen:2014gka} for each term in the sum of~\eref{diffwidth},
\bea
\label{eq:absdiffw}
\Pi_{nm}^{ij} &=&
\int \left| \frac{d\Gamma_{nm}^{ij}}{d\mathcal{P}} \right| d\mathcal{P} ,
\eea
where the $\Pi_{nm}^{ij}$ are positive definite even in the case of CP violation.~These integrated magnitudes contain information not only about the total phase space contribution of each combination of operators, but also about the differences in shape of the differential spectra.~It is for this reason that one can have non-zero values even for combinations of operators which lead to $CP$ violation.

We show in~\fref{absnorm1} the $\Pi_{nm}^{ij}$ for the $2e2\mu$ final state (with similar numbers for $4e,~4\mu$) assuming a resonance mass of $m_{\varphi} = 18.4$~GeV which corresponds approximately to the lightest predicted mass in~\cite{Chen:2016jxd}.~We have also imposed cuts on the dilepton invariant mass as well as lepton transverse momentum and rapidity of $M_{\ell\ell} > 0.1$~GeV,~$p_{T\ell} > 2$~GeV, and $\eta_{\ell} < 2.4$ while normalizing to the $|\mathcal{A}_1^{\Upsilon\Upsilon}|^2$ term, thus giving one for that entry.~We see by examining the diagonal terms that the largest integrated magnitudes are for the $\Upsilon\gamma$ operators.~This is due to a combination of the fact that in these cases both gauge bosons are kinematically allowed to be close to on-shell, as well as the larger coupling of photons to leptons relative to the $Z$ and $\Upsilon$.~The next largest contributions come from the $|\mathcal{A}_1^{\Upsilon\Upsilon}|^2$ term where the second $\Upsilon$ is kinematically restricted to be off-shell by around $500$~MeV which is much larger than the $\Upsilon$ width.

As expected due to the large $Z$ boson mass, operators involving the $Z$ boson are very suppressed relative to the others.~Thus any observable effects due to these operators would require extremely large couplings in comparison to $\mathcal{A}_n^{\Upsilon\Upsilon}$ and $\mathcal{A}_n^{\Upsilon\gamma}$.~However, since we seek to be as agnostic as possible about any potential resonance and so little is known \emph{a priori} about the possible form factors, we include them in our analysis.~Furthermore, since we are not using rate information in the pure hypothesis tests conducted below, we can use only \emph{shape} information to disfavor one operator for another independently of the size of the couplings.~Of course the ultimate sensitivity and precision with which the form factors can be measured will depend on their numerical value (and potential momentum dependence).~We also see that interference effects, especially in the case of $\Upsilon\Upsilon$ and $\Upsilon\gamma$ operators can be relevant and opens the possibility of using parameter extraction methods~\cite{Chen:2012jy,Chen:2013ejz,Khachatryan:2014kca} to directly measure phases or possibly $CP$ violation.~Though we don't do so here, it would be interesting to explore these potential interference effects should a resonance be observed.
\begin{figure}[t]
\includegraphics[width=.465\textwidth]{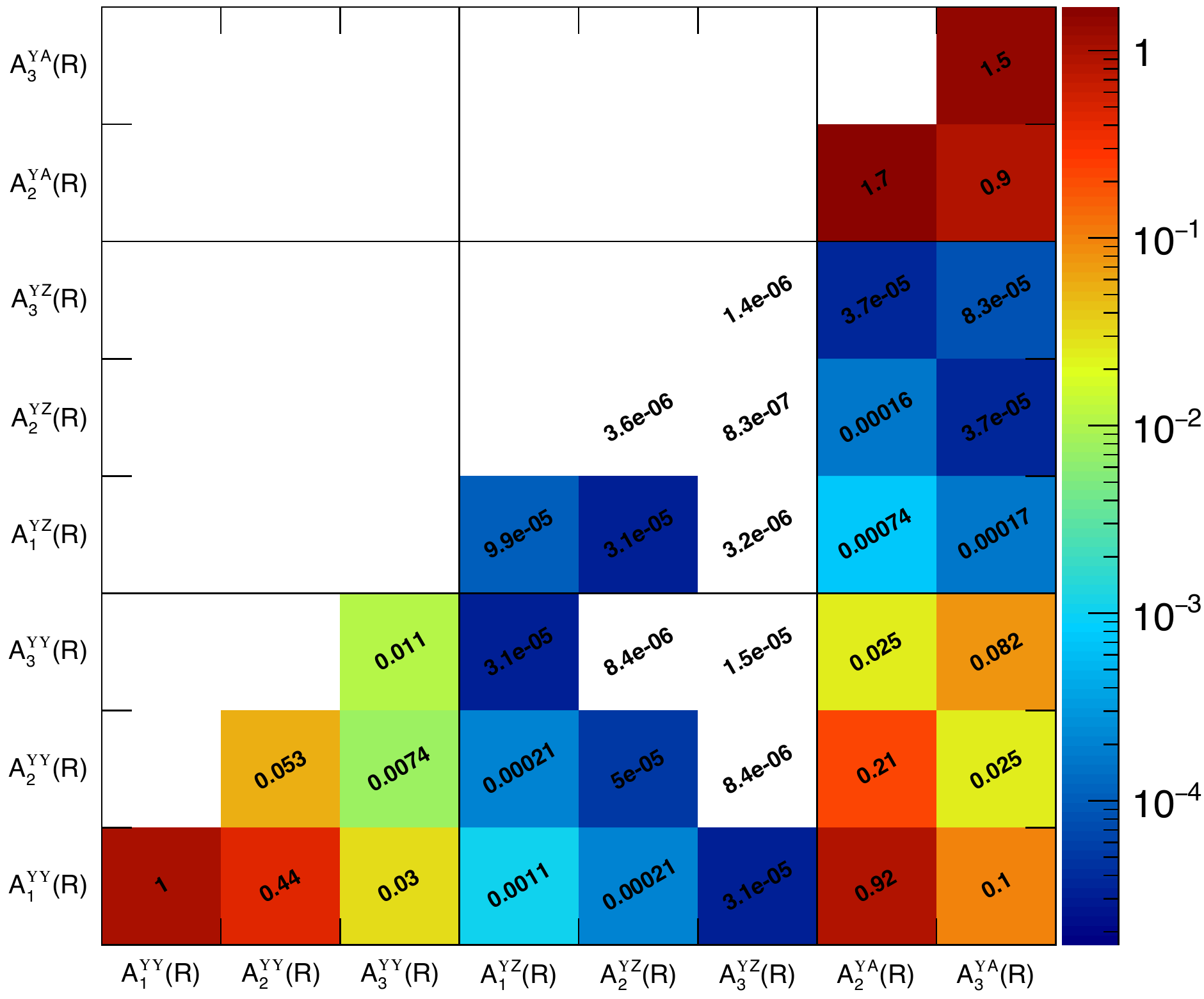}
\caption{Total integrated magnitudes, $\Pi^{ij}_{nm}$, defined in~\eref{absdiffw}, which correspond to the pairs of couplings $\mathcal{A}^i_n \mathcal{A}^{j\ast}_m$ defined in~\eref{LM} and~\eref{LVV}.~We have normalized to the $|\mathcal{A}^{\Upsilon\Upsilon}_1|^2$ diagonal term and taken $m_{\varphi} = 18.4$~GeV corresponding approximately to the lightest predicted mass in~\cite{Chen:2016jxd}.~We have also imposed cuts on the dilepton invariant mass of $M_{\ell\ell} > 0.1$ as well as lepton $p_T$ and rapidity of $p_{T\ell} > 2$~GeV, and $\eta_{\ell} < 2.4$ for the $2e2\mu$ final state (with similar numbers for $4e,~4\mu$).}
\label{fig:absnorm1}
\end{figure}

\subsection{One dimensional projections} \label{sec:proj}

Since we do not use rate information in our MEM analysis, the discriminating power of the $\varphi \to 4\ell$ channel comes from differences in shape of the differential spectra arising from the operators in~\eref{LM} and~\eref{LVV}.~To gain intuition for this we examine various one dimensional projections.~As discussed above, in the $\varphi$ center of mass frame (or four lepton) there are two invariant masses formed out of the four vectors of the di-lepton pairs as well as five angular variables~\cite{Chen:2013ejz}.
\begin{figure}[tbh]
\begin{center}
\includegraphics[scale=.47]{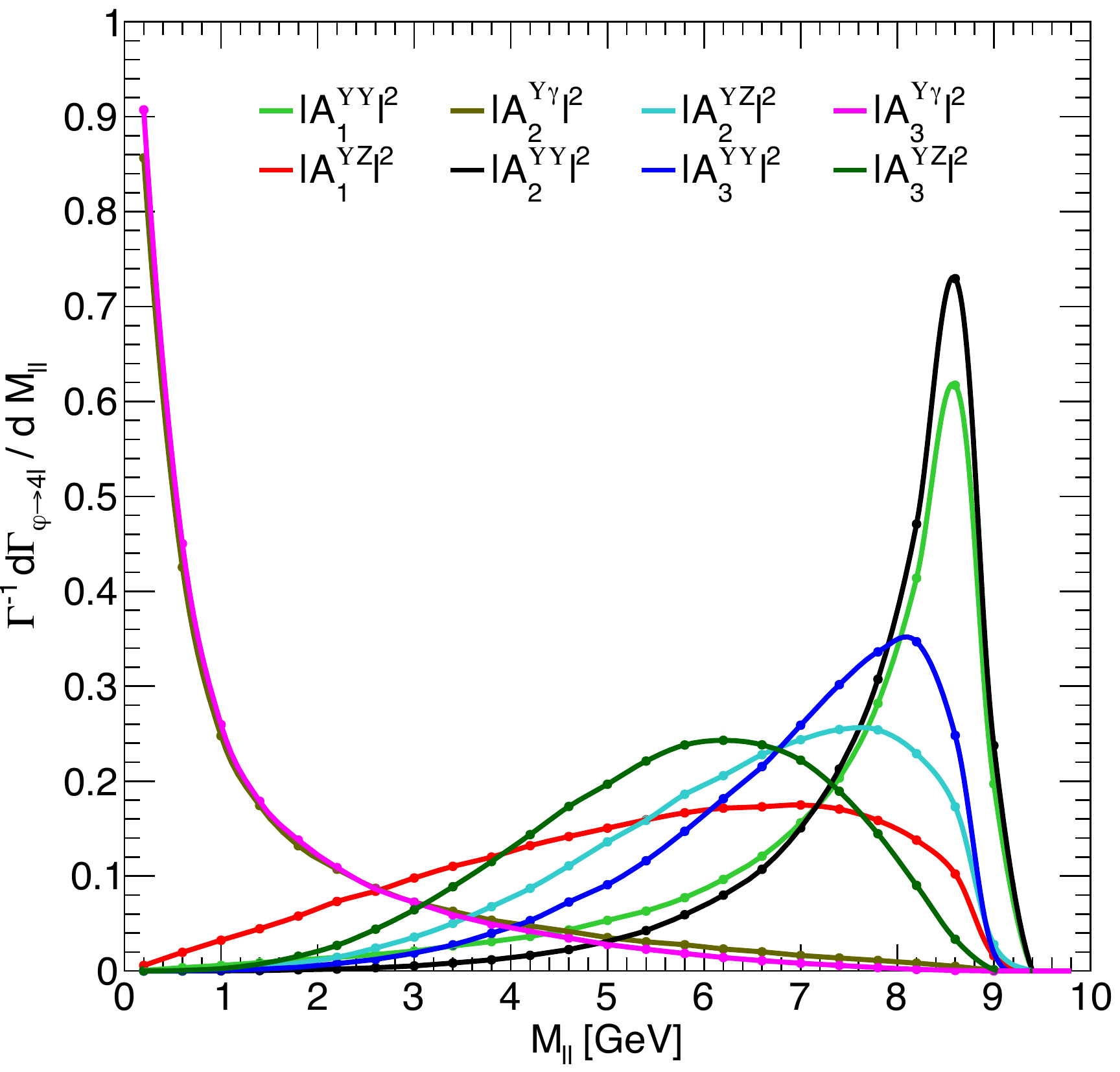}
\caption{Normalized differential spectra of the dilepton invariant mass $M_{\ell\ell}$ for the eight `diagonal' terms in~\eref{diffwidth} with operators defined in~\eref{LM} and~\eref{LVV}.~We show the $2e2\mu$ final state after integrating over the phase space in~\fref{absnorm1}.
}
\label{fig:M2}
\end{center}
\end{figure}

The three most discriminating variables are the `off-shell' dilepton mass $M_{\ell\ell}$,~the polar angle $\theta_{\ell\ell}$ of the lepton momentum in the rest frame of the off-shell dilepton pair,~and the azimuthal angle $\Phi$ between the decay planes of the lepton pairs.~These are shown in~\fref{M2} and~\fref{angles} again assuming a resonance mass $m_{\varphi} = 18.4$~GeV.~For each projection we have integrated the other kinematic variables over the phase space~\cite{Chen:2013ejz} defined above and in~\fref{absnorm1}.~We show distributions for the $|\mathcal{A}_{1}^{\Upsilon\Upsilon}|^2$ (black),~$|\mathcal{A}_{2}^{\Upsilon\Upsilon}|^2$ (blue),~$|\mathcal{A}_{3}^{\Upsilon\Upsilon}|^2$ (turqoise),~$|\mathcal{A}_{1}^{\Upsilon Z}|^2$ (red),~$|\mathcal{A}_{2}^{\Upsilon Z}|^2$ (gold),~and~$|\mathcal{A}_{2}^{\Upsilon \gamma}|^2$ (green) `diagonal' terms in~\eref{diffwidth}.~There is also a polar angle for the lepton coming from the on-shell $\Upsilon$ decay which has a very similar distribution to $\theta_{\ell\ell}$ and also aids in discrimination.~From these distributions it is clear that the differential spectra contain information about the pole structure of the operators as well as the $CP$ and tensor properties.

Of course the fully differential decay width used in our MEM analysis contains more information than these projections and includes all correlations.~Furthermore, utilizing all decay variables aids in discrimination against  backgrounds regardless of if they help in signal discrimination.~Strong momentum effects in the $\mathcal{A}_{n}^{i}$ form factors may distort these spectra, but unless there are additional poles they should not drastically change the overall shape of these distributions, particularly the angular ones.
\begin{figure}[tbh]
\begin{center}
\includegraphics[scale=.44]{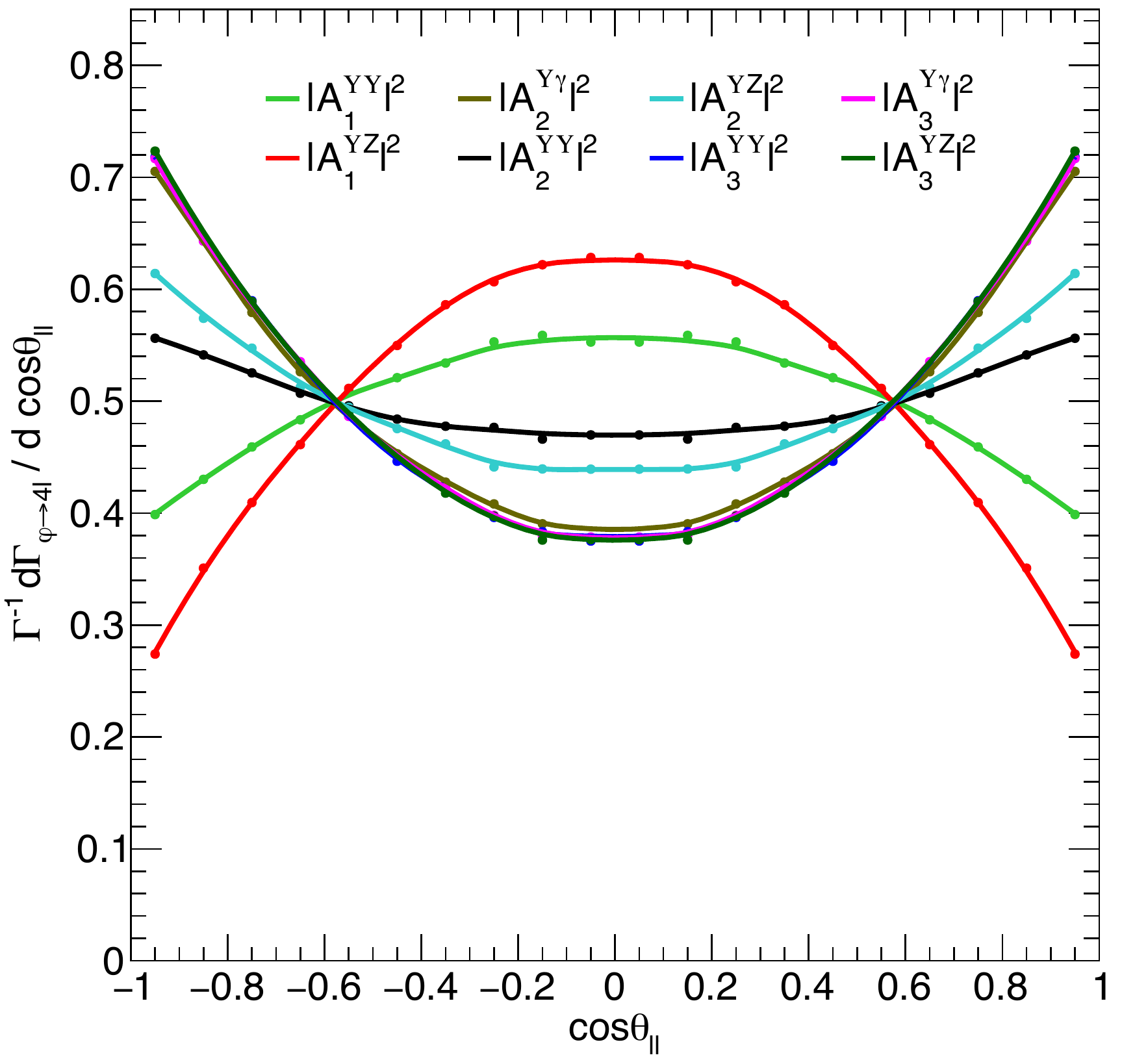}\\
\includegraphics[scale=.435]{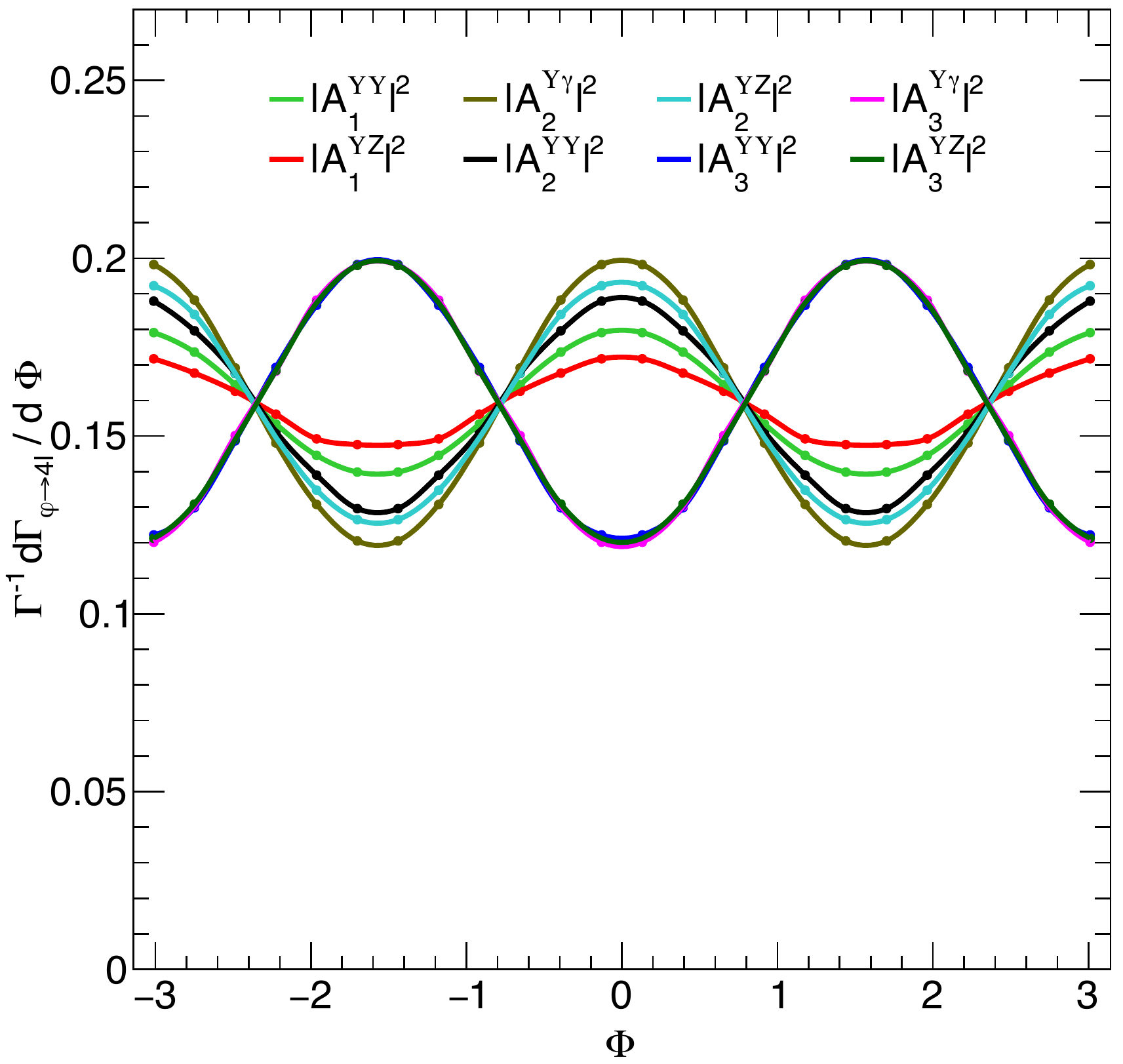}
\caption{{\bf Top:}~Normalized differential spectra for the polar angle $\theta_{\ell\ell}$ of the lepton momentum in the rest frame of the dilepton pair for the same operators and final state as in~\fref{M2}.~{\bf Bottom:}~Same as top, but for the azimuthal angle $\Phi$ between the decay planes of the two lepton pairs.
}
\label{fig:angles}
\end{center}
\end{figure}

\subsection{Opportunities for $CP$ violation} \label{sec:cpv}

Here we briefly discuss some of the potential sources of $CP$ violation which might be present in these decays.~The possibility of additional poles entering in the form factors $\mathcal{A}_n^{i}$ and generating an imaginary component which can act as a `strong phase' allows for additional sources of $CP$ violation apart from those due to interference between $CP$ even and $CP$ odd operators in~\eref{LVV} with real coefficients.~With strong phases present in the $\mathcal{A}_n^{i}$ form factors we can construct $CP$ violating observables without requiring four reconstructed four vectors and which can show up for example as an asymmetry in the $\theta_{\ell\ell}$ distribution~\cite{Chen:2014ona,Li:2015kxc,Chen:2017plj}.~These strong phase $CP$ violating effects are in contrast to more standard `weak phase' $CP$ violation which occurs when there are $CP$ odd and $CP$ even operators (such as $\mathcal{A}_2^{i}$ and $\mathcal{A}_3^{i}$) with the couplings purely real.~This type of $CP$ violation does require constructing $CP$ violating observables via triple products from four reconstructed four vectors and, as is well known from studies of Higgs boson decays to four leptons~\cite{Cao:2009ah}, manifests as a phase shift in the $\Phi$ distribution.

Of course the presence of poles may also introduce strong momentum effects.~A more detailed exploration of all of these possible effects is beyond the scope of this study, should a beautiful tetraquark or some other resonance near the di$-\Upsilon$ threshold be discovered, it would be interesting to investigate these potential $CP$ violating effects and to what extent they can be distinguished.

\section{MEM Analysis of $\varphi \to 4\ell$ decays} \label{sec:LHC}

Here we present results of our MEM analysis to assess the ability of the four lepton channel to probe the nature of a resonance around the di$-\Upsilon$ threshold.~In particular we examine the sensitivity for distinguishing the different operators in~\eref{LM} and~\eref{LVV} in order to establish its CP and tensor properties.~Details of our analysis procedure based on utilizing the normalized fully differential $\varphi \to 4\ell$ decay width to construct the likelihood can be found in~\cite{Stolarski:2012ps,Chen:2013ejz,Falkowski:2014ffa,Chen:2015rha,Chen:2016ofc}.~To generate events we have implemented the tensor structures in~\eref{LM} and~\eref{LVV} into the Feynrules/Madgraph framework~\cite{Alwall:2014hca,Christensen:2008py}.

Motivated by the interesting possibility of a bottom tetraquark bound state~\cite{Berezhnoy:2011xn,Du:2012wp,Chen:2015moa,Karliner:2016zzc,Chen:2016jxd,Bai:2016int,Wu:2016vtq,Richard:2017vry,Wang:2017jtz}, for most of our MEM analysis we take the resonance mass to be $m_{\varphi} = 18.4$~GeV corresponding approximately to the lightest predicted mass in~\cite{Chen:2016jxd} and assume experimental cuts on the phase space of the final state leptons of $M_{\ell\ell} > 0.1$~GeV,~$p_{T\ell} > 2$~GeV, and $\eta_\ell < 2.4$ accounting for the relevant selection efficiencies.~To attempt to model the background we include the leading order $q\bar{q}\to 4\ell$ process also computed analytically and validated in~\cite{Chen:2012jy,Chen:2013ejz}.~There are NLO corrections~\cite{Frederix:2011ss,Biedermann:2016lvg} and unknown, but potentially large experimental backgrounds.~However, judicious choice of experimental cuts~\cite{Khachatryan:2016ydm}, including requiring an on-shell $\Upsilon$ and a narrow window around the $\varphi$ mass, should greatly reduce the four lepton background.~We therefore simply float the background fraction~\cite{Chen:2013ejz,Chen:2014pia} and assume during event generation that it is $\sim 50\%$ at the resonance mass.

We then perform a simplified study to assess how the ability to distinguish between the operators in~\eref{LM} and~\eref{LVV} depends on how close the resonance mass is to the di$-\Upsilon$ threshold and whether it is above or below.

\subsection{Hypothesis testing} \label{sec:hyptest}

We use the hypothesis testing techniques developed in~\cite{DeRujula:2010ys} and utilized in~\cite{Stolarski:2012ps,Falkowski:2014ffa,Chen:2016ofc} to construct a test statistic that measures the separation power between pairs of operators in~\eref{LM} and~\eref{LVV}.~We do this by constructing the ratio between two likelihoods assuming only one of the operators dominates at a time in each likelihood.~Pseudoexperiments are then conducted to obtain a distribution of these likelihood ratios.~This is first done assuming one operator as the `true' hypothesis and then repeated assuming the second operator corresponds to the true hypothesis.~In each case, a distribution of likelihood ratios is obtained after conducting a large set of psuedoexperiments.~The overlap (or lack thereof) between these two distributions can then be converted into a measure of the ability to discriminate between the two models.~We follow closely the procedure in~\cite{Stolarski:2012ps}, but present our results in terms of $p\,$-values instead of $\sigma$.

\subsection{Probing a beautiful tetraquark} \label{sec:4b}

Motivated by the possible existence~\cite{CMSpub} of a four bottom quark bound state at $m_\varphi = 18.4$~GeV, we show in~\fref{hyptest} the probability of mistaking the $\mathcal{A}_{1}^{\Upsilon\Upsilon}$ operator for $\mathcal{A}_{2}^{\Upsilon\Upsilon}$ (black),~$\mathcal{A}_{3}^{\Upsilon\Upsilon}$ (blue),~$\mathcal{A}_{1}^{\Upsilon Z}$ (red),~$\mathcal{A}_{2}^{\Upsilon Z}$ (turqoise),~$\mathcal{A}_{2}^{\Upsilon \gamma}$ (forest green),~$\mathcal{A}_{3}^{\Upsilon Z}$ (dark green),~$\mathcal{A}_{3}^{\Upsilon \gamma}$ (pink) or \emph{vice versa} as a function of the number of four lepton signal events for the operators defined in~\eref{LM} and~\eref{LVV}.~We see that $\varphi \to \Upsilon \,\ell^+\ell^- \to 4\ell$ decays should be able to discriminate at $\sim95\%$ confidence between the $\mathcal{A}_{1}^{\Upsilon\Upsilon}$ operator and the other possibilities with as few as 10 events in the case of $\mathcal{A}_{2}^{\Upsilon\gamma}$ while $\sim 150$ signal events will be needed to distinguish it from $\mathcal{A}_{2}^{\Upsilon\Upsilon}$ which has the same $CP$ and pole structure.~Since there is not a clear prediction for the production cross section, it is difficult to give robust results in terms of luminosity.~However, assuming a $4b$ tetraquark and a gluon initiated process, the results from~\cite{Eichten:2017ual} estimate a $\varphi \to 4\ell$ production cross section of $\mathcal{O}(1)\,\rm{fb^{-1}}$ at a 13~TeV LHC.~In this case our results indicate that $\mathcal{O}(10-150)\,\rm{fb^{-1}}$ of luminosity would be needed to probe its bottom tetraquark nature which may be achievable perhaps not long after a discovery. 
\begin{figure}[tbh]
\begin{center}
\includegraphics[scale=.45]{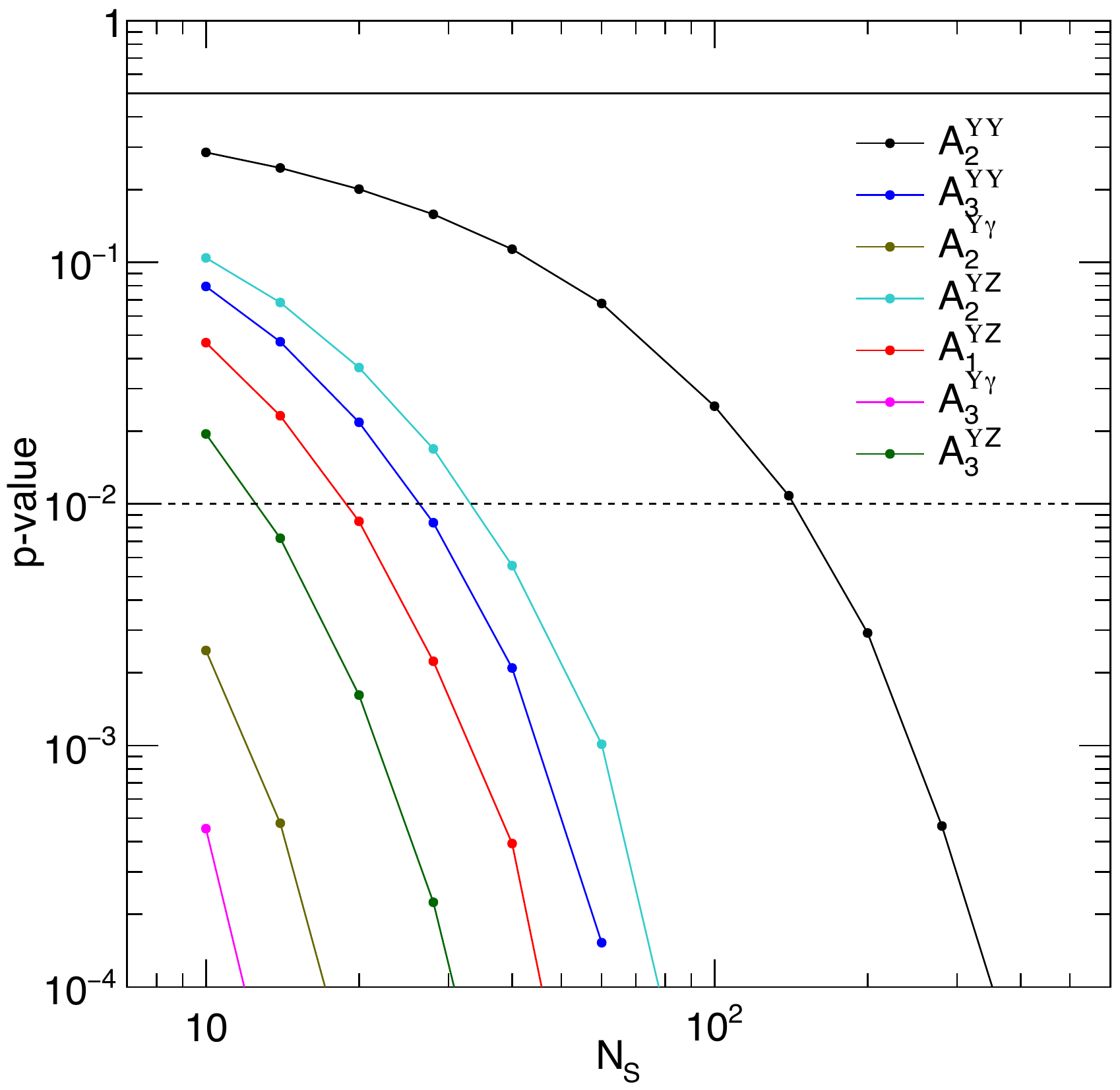}
\caption{Probability  (in $p$-values) of mistaking the $\mathcal{A}_{1}^{\Upsilon\Upsilon}$ operator (as defined in~\eref{LM} and~\eref{LVV}) for $\mathcal{A}_{2}^{\Upsilon\Upsilon}$ (black),~$\mathcal{A}_{3}^{\Upsilon\Upsilon}$ (blue),~$\mathcal{A}_{1}^{\Upsilon Z}$ (red),~$\mathcal{A}_{2}^{\Upsilon Z}$ (turqoise),~ $\mathcal{A}_{2}^{\Upsilon \gamma}$ (forest green),~$\mathcal{A}_{3}^{\Upsilon Z}$ (dark green),~$\mathcal{A}_{3}^{\Upsilon \gamma}$ (pink) or \emph{vice versa} versus the number of four lepton signal events ($2e2\mu + 4e + 4\mu$) for a resonance mass of $m_\varphi = 18.4$~GeV.~We have floated the background fraction~\cite{Chen:2013ejz,Chen:2014pia} and assume during event generation that it is $\sim 50\%$ at the resonance mass.}
\label{fig:hyptest}
\end{center}
\end{figure}

\subsection{Sensitivity around the di$-\Upsilon$ threshold} \label{sec:thresh}

We also examine how the ability to distinguish between the operators in~\eref{LM} and~\eref{LVV} depends on how close to the di$-\Upsilon$ threshold the resonance mass is.~For this we again perform hypothesis tests for a range of resonance masses around the di$-\Upsilon$ threshold.~In~\fref{hyptest2} we show the number of four lepton signal events needed to distinguish the $\mathcal{A}_{1}^{\Upsilon\Upsilon}$ operator from $\mathcal{A}_{2}^{\Upsilon\Upsilon}$ (black),~$\mathcal{A}_{3}^{\Upsilon\Upsilon}$ (blue),~$\mathcal{A}_{1}^{\Upsilon Z}$ (red),~$\mathcal{A}_{2}^{\Upsilon Z}$ (turqoise),~$\mathcal{A}_{2}^{\Upsilon \gamma}$ (forest green),~$\mathcal{A}_{3}^{\Upsilon Z}$ (dark green),~$\mathcal{A}_{3}^{\Upsilon \gamma}$ (pink) or \emph{vice versa} with $95\%$ confidence \emph{versus} the four lepton resonance mass.~The points correspond to the values for the resonance mass of $m_\varphi = 18.4, 18.8, 18.9, 18.92\,(\rm{threshold}), 18.94, 19.0, 19.3$~GeV.~In this case we have neglected backgrounds and assumed 5 signal events for the $2e2\mu$ channel and 5 events for $4e + 4\mu$ channel before combining into one likelihood~\cite{Chen:2015rha,Chen:2016ofc} formed out of the $\varphi \to 4\ell$ fully differential decay width.
\begin{figure}[tbh]
\begin{center}
\includegraphics[scale=.45]{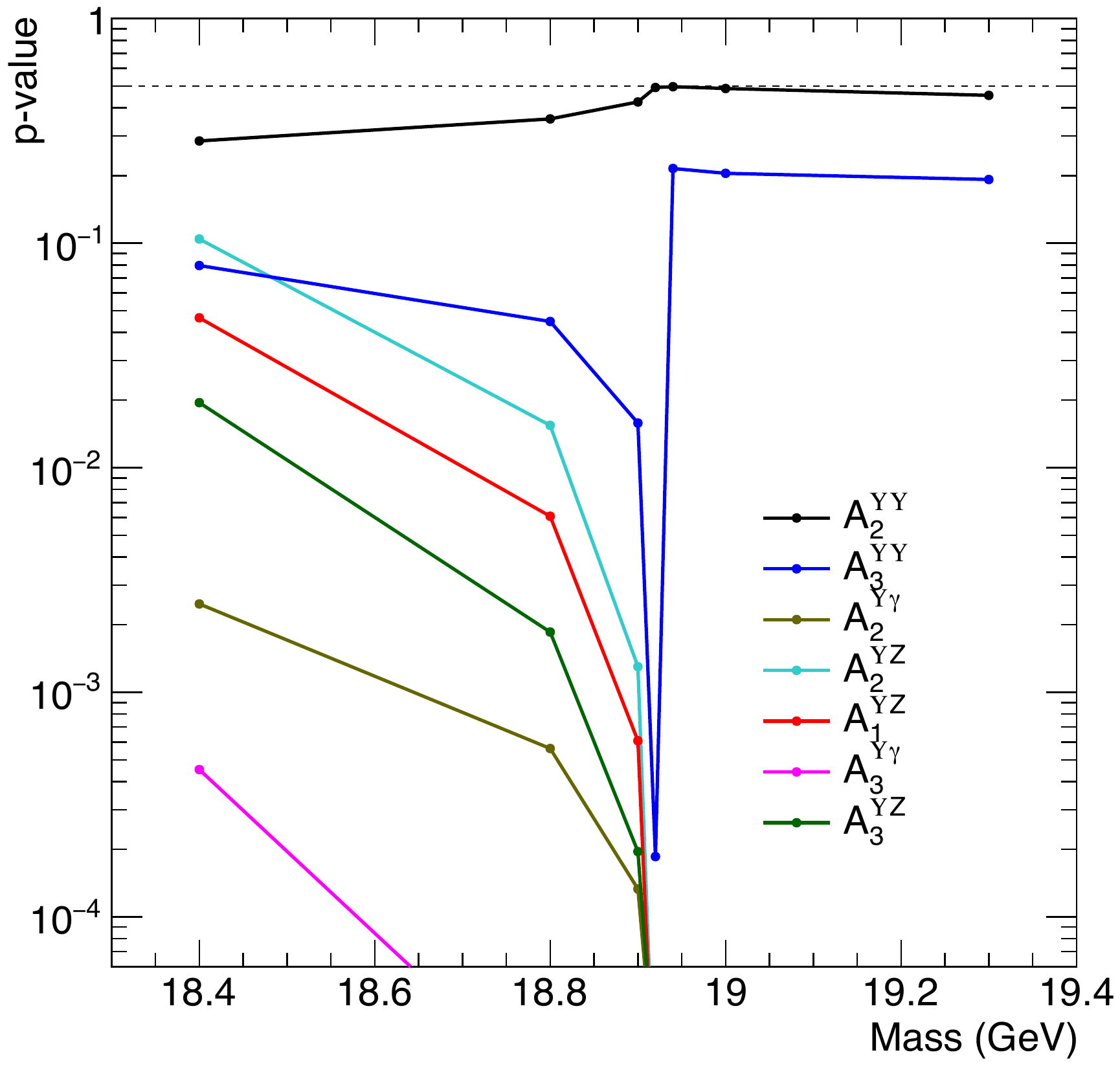}
\caption{The probability (in $p$-values) of mistaking the $\mathcal{A}_{1}^{\Upsilon\Upsilon}$ operator (as defined in~\eref{LM} and~\eref{LVV}) for $\mathcal{A}_{2}^{\Upsilon\Upsilon}$ (black),~$\mathcal{A}_{3}^{\Upsilon\Upsilon}$ (blue),~$\mathcal{A}_{1}^{\Upsilon Z}$ (red),~$\mathcal{A}_{2}^{\Upsilon Z}$ (turqoise),~$\mathcal{A}_{2}^{\Upsilon \gamma}$ (forest green),~$\mathcal{A}_{3}^{\Upsilon Z}$ (dark green),~$\mathcal{A}_{3}^{\Upsilon \gamma}$ (pink) or \emph{vice versa} versus four lepton resonance mass.~The points correspond to four lepton resonance masses of $m_\varphi = 18.4, 18.8, 18.9, 18.92\,(\rm{di}-\Upsilon~\rm{threshold}), 18.94, 19.0, 19.3$\,GeV.~We have neglected backgrounds and assumed 5 signal events for the $2e2\mu$ channel and 5 events for $4e + 4\mu$ channel before combining into one likelihood~\cite{Chen:2015rha,Chen:2016ofc} for the $\varphi \to 4\ell$ signal.}
\label{fig:hyptest2}
\end{center}
\end{figure}

We see the ability to distinguish between the different operators in~\eref{LM} and~\eref{LVV} depends quite strongly on the exact resonance mass.~In general we see that the operator most difficult to distinguish from $\mathcal{A}_{1}^{\Upsilon\Upsilon}$, is $\mathcal{A}_{2}^{\Upsilon\Upsilon}$ which has the same $CP$ and pole structure.~Furthermore, we see that the nearer to threshold the more difficult it is to distinguish these two operators.~Interestingly, we see that distinguishing between $\mathcal{A}_{1}^{\Upsilon\Upsilon}$ and the $CP$ odd $\mathcal{A}_{3}^{\Upsilon\Upsilon}$ operator is much easier and becomes more so the closer to the di-$\Upsilon$ threshold we get.~This is due to a combination of the $\Phi$ distribution (see~\fref{angles}) as well as the di-lepton invariant mass distribution ($M_{\ell\ell}$), which in general is not peaked at the same mass and has a different slope than the $M_{\ell\ell}$ distribution in $\mathcal{A}_{1}^{\Upsilon\Upsilon}$ and $\mathcal{A}_{2}^{\Upsilon\Upsilon}$.~In other words, the $M_{\ell\ell}$ distribution also contains information about the $CP$ properties of the $\varphi\Upsilon\Upsilon$ distribution (see~\fref{M2}) which of course has also been noted in SM-like Higgs boson decays to four leptons~\cite{Boughezal:2012tz}.~In general we see that, depending on the nature and mass of a putative resonance around the di-$\Upsilon$ threshold, just a few handful of signal events may be sufficient for deciphering its $CP$ and tensor properties in the four lepton channel.

The strong dependence on the exact resonance implies that detector resolution effects may play an important role and need to be included for a precise quantification of the sensitivity.~While we have not done so for this preliminary study, including these detector effects can be done with the framework developed in~\cite{Chen:2014pia,Chen:2014hqs} which we leave to future work should a four lepton resonance be discovered near the di-$\Upsilon$ threshold.

\subsection{Opportunities for parameter extraction} \label{sec:fits}

One of our primary assumptions in this analysis is that only one operator from~\eref{LM} and~\eref{LVV} dominates at a time.~We have also neglected any momentum dependence in the form factors as well as parton showering and jets.~These assumptions of course may not be true in reality and therefore an analysis which can extract multiple form factors simultaneously with possible momentum dependence is also important.~Furthermore, as~\fref{absnorm1} shows, even if one operator dominates, interference effects may be relevant which would allow us to probe a multitude of effects.~As more four lepton events are collected, utilizing multi-parameter fitting methods~\cite{Chen:2013ejz,Chen:2014pia} to extract the various form factors may become possible and also opens the possibility of probing $CP$ violating effects.~However, this requires a more careful examination of backgrounds and detector resolution effects so we leave this to interesting possibilities which can be explored should a four lepton resonance around $\sim 18 - 19$~GeV be observed.

\section{Summary and conclusions} \label{sec:const}

Motivated by recent studies~\cite{Berezhnoy:2011xn,Du:2012wp,Chen:2015moa,Karliner:2016zzc,Chen:2016jxd,Bai:2016int,Wu:2016vtq,Richard:2017vry,Wang:2017jtz,Eichten:2017ual} which indicate that a QCD state composed of four bottom quarks may be present around around the di$-\Upsilon$ threshold in the range $\sim 18 - 19$~GeV and potentially observable in its decays to four leptons, we have explored the possibility of using the \emph{normalized} fully differential decay width to four leptons to probe its underlying nature and in particular its $CP$ and tensor properties.~Further motivated by the exciting possibility of a tetraquark bound state explored in some of these studies~\cite{Berezhnoy:2011xn,Du:2012wp,Chen:2015moa,Karliner:2016zzc,Chen:2016jxd,Bai:2016int,Wu:2016vtq,Richard:2017vry,Wang:2017jtz}, we emphasize in our analysis a spin-0 resonance $\sim 500$~MeV \emph{below} the di-$\Upsilon$ threshold, but also examine  resonance masses just below, at, and above the di-$\Upsilon$ threshold.

Assuming the putative resonance is a spin-0 boson, we have performed a simplified matrix element method analysis based on hypothesis testing and assuming its decays are dominated by one effective operator at a time.~We find excellent prospects at the LHC once $\mathcal{O}(10-100)$ signal events are collected indicating that perhaps its underlying nature, including $CP$ properties, can be directly probed.~This may in principle be possible not long after discovery depending on (unknown) production cross sections and the exact resonance mass.~We have also briefly discussed various potential $CP$ violating effects which might be present in these decays, but we leave a study of this as well as spin-1 and spin-2 resonances to ongoing~\cite{followup2} and future work.

While this study is motivated by possible four bottom quark QCD states near the di$-\Upsilon$ threshold, our analysis methods easily generalize to any potential light scalar, including other multi-quark states, which can decay to four lepton final states at the LHC or other colliders.

~\\
\noindent
{\bf Acknowledgments:}~We especially thank Ciaran Hughes and Clara Peset for illuminating discussions and for helping reduce the author's ignorance on quark bound states.~We also thank Maria Spiropulu for providing us with computing resources necessary to complete this study as well as Adam Falkowski,~Jose Santiago,~and Daniel Stolarski for useful comments and discussions.~The work of R.V.M.~is supported by MINECO, FPA 2016-78220-C3-1-P, FPA 2013-47836-C3-2/3-P (including ERDF), and the Juan de la Cierva program,~as well as by Junta de Andalucia Project FQM-101.~R.V.M.~would also like to thank the Mainz Institute for Theoretical Physics (MITP) for its hospitality and partial support during the completion of this work as well as Fermilab National Accelerator Laboratory and Northwestern University for their hospitality.
%

\bibliographystyle{apsrev}
\bibliography{references}

\end{document}